\documentclass[10pt,aps,physrev,raggedbottom,longbibliography,nobalancelastpage,reprint,citeautoscript,letterpaper,superscriptaddress]{revtex4-2}
\usepackage[T1]{fontenc}
\usepackage{bm}
\usepackage{braket}
\usepackage{multirow}
\usepackage{bbold}
\usepackage{graphicx}
\usepackage{subfigure}
\usepackage{amssymb}
\usepackage{amsmath}
\usepackage{microtype}
\usepackage{lmodern}
\usepackage[scaled=0.875]{helvet}

\usepackage[dvipsnames]{xcolor}

\definecolor{codegreen}{rgb}{0,0.6,0}
\definecolor{codegray}{rgb}{0.5,0.5,0.5}
\definecolor{codepurple}{rgb}{0.58,0,0.82}
\definecolor{backcolour}{rgb}{0.95,0.95,0.92}

\usepackage[bookmarks=false,colorlinks]{hyperref}
\usepackage[all]{hypcap} 
\hypersetup{linkcolor=magenta,citecolor=MidnightBlue,filecolor=Plum,urlcolor=MidnightBlue}

\usepackage{listings}
\lstdefinestyle{mystyle}{
    backgroundcolor=\color{backcolour},   
    commentstyle=\color{codegreen},
    keywordstyle=\color{magenta},
    numberstyle=\tiny\color{codegray},
    stringstyle=\color{codepurple},
    basicstyle=\ttfamily\footnotesize,
    breakatwhitespace=false,         
    breaklines=true,                  
    captionpos=b,                    
    keepspaces=true,                  
    numbers=left,                    
    numbersep=5pt,                  
    showspaces=false,                
    showstringspaces=false,
    showtabs=false,                  
    tabsize=2
}
\usepackage{ccaption}
\usepackage{ragged2e}
\captionnamefont{\small \sffamily \bfseries }
\captiontitlefont{\small }
\captiondelim{~\textbar~}
\captionstyle{\justifying}

\newcommand{\Ham}{\mathbf{H}}
\newcommand{\Smat}{\mathbf{S}}
\newcommand{\irr}[2]{\ensuremath{#1\!\otimes\!#2}}
\newcommand{\Dmat}[1]{\mathbf{D}^{(#1)}}
\newcommand{\CG}[1]{C^{(#1)}}



\begin{document}
\preprint{APS/123-QED}

\title{Towards a universal model for spin-orbit coupled Wannier Hamiltonians}

\author{Alexander C.\ Tyner}

\affiliation{Tailwater Informatics, San Diego, CA, USA}

\date{\today}

\begin{abstract} 
While machine learning interatomic potentials (MLiPs) have matured to revolutionize material science, deep learning models for electronic structure are just beginning to emerge and restricted, almost exclusively, to non-orthogonal basis Hamiltonians. We introduce G(Wa)NN, the first deep-learning model capable of generating the electronic Hamiltonian of solid-state systems in an orthogonal Wannier basis. G(Wa)NN is trained on an unprecedented, diverse dataset of more than 111K Wannier Hamiltonians (150M+ hopping matrices) spanning 69 elements. The combination of optimized inference and linear-scaling methods for orthogonal Hamiltonians unlock transport simulations at massive scales (10K+ atoms). Crucially, the framework supports local finetuning, allowing users to adapt the base model to custom Wannier Hamiltonian datasets. To seamlessly translate these predictions into physical observables, we introduce Tailwater, a Python package providing an API interface to G(Wa)NN alongside a high-performance post-processing library. Tailwater enables automated projection of the predicted Hamiltonian into an arbitrary low-energy subspace—directly mirroring familiar Wannier90 workflows—and includes a suite of Kernel Polynomial Method (KPM) functions that exploit the orthogonal basis to achieve strict linear scaling for spectral observables. The Tailwater ecosystem, with the G(Wa)NN model at its core, aims to help bridge the gap between deep learning and macro-scale quantum transport simulations.
\end{abstract}

\maketitle

\section{Introduction} 
As electronic devices reach ultra-scaled dimensions and exploit quantum-mechanical phenomena, keeping pace with Moore's law demands radical materials innovation\cite{Gall:2022,chen2020topological,zou2021review,gall2021materials,moon2023materials}. Artificial intelligence has emerged as a cornerstone of this effort, most notably through machine-learned interatomic potentials (MLiPs)\cite{batatia2025foundation}. Architectures such as MACE\cite{mace}, CHGNet\cite{deng_2023_chgnet}, EquiformerV2\cite{equiformer}, and M3GNet\cite{chen2022universal} are trained on millions of first-principles calculations\cite{barros2026open} to map potential energy surfaces with DFT-level accuracy at a fraction of the computational cost. This paradigm shift has enabled macroscopic simulations, transforming workflows across drug discovery, carbon capture, and energy storage.
\par
\begin{figure*}
    \centering
    \includegraphics[width=18cm]{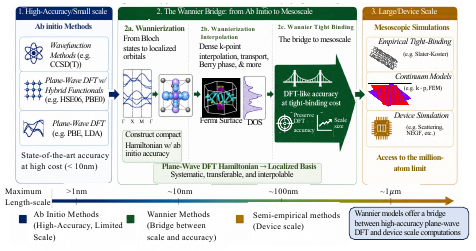}
    \caption{\textbf{Bridging the computational gap between plane-wave DFT and mesoscopic computations with Wannier Hamiltonians:} Traditional first-principles methods (top left), such as plane-wave DFT (hybrid-functionals in particular) and advanced wave function theories, offer high electronic structure accuracy but generally remain confined to the sub-nanometer scale due to steep $O(N^3)$ computational scaling bottlenecks. Conversely, empirical tight-binding and continuum models (right) scale to mesoscopic dimensions ($\sim1\mu\text{m}$) but sacrifice explicit quantum fidelity and predictive accuracy. Wannier-Tight Binding (WTB) models serve as a critical bridge to the mesoscale. By utilizing a naturally orthogonal, maximally localized basis, the Wannier representation delivers a mix of speed and accuracy—maintaining near first-principles predictive fidelity while operating at a tight-binding computational cost to unlock simulations across extended length scales.}
    \label{fig:Fig0}
\end{figure*}
\par 
Despite the maturity of MLiPs, deep learning frameworks for generating electronic structure remain in their infancy. This lag stems from data scarcity, the challenge of capturing long-range interactions, and variations in basis-set choices. Yet, foundational models for electronic structure are vital to accelerate electronic materials discovery. This need is starkly apparent in the search for next-generation interconnects in integrated circuits\cite{kiani2025searching,Tyner2026,kiani2025shrinking,kim2025future}. While MLiPs can assess structural manufacturability, optimizing an interconnect requires evaluating its electrical conductance—a metric fundamentally tied to electronic structure. Currently, large-scale screening of massive virtual databases like GNOME\cite{merchant2023scaling} remains out of reach because conventional first-principles transport calculations cannot handle the necessary system sizes. A foundational electronic structure model capable of linear scaling could rapidly screen these repositories.

\par 
Pioneering frameworks such as Uni-HamGNN\cite{zhong2026universal}, DeepH-E3\cite{gong2023general}, and MACE-H\cite{qian2026equivariant} have made significant strides by training graph neural networks (GNNs) on Hamiltonians expressed in non-orthogonal Linear Combination of Atomic Orbitals (LCAO) bases. LCAO representations provide a fixed spatial basis that naturally limits gauge degrees of freedom, standardizing the target tensor for covariant machine learning models. However, mapping the electronic structure to an LCAO basis introduces a downstream consequence: it yields a non-orthogonal Hamiltonian ($H$) governed by the generalized eigenvalue problem $H\Psi=SE\Psi$, where $S$ is the overlap matrix. While GNNs can accurately predict $S$, quantum transport codes—particularly those utilizing the Kernel Polynomial Method (KPM) for $O(N)$ scaling\cite{kpm}—frequently require an orthogonal basis. Transforming the system via $H_{\text{orth}}=S^{-1/2}HS^{-1/2}$ destroys the sparsity of the matrices, yielding a dense $H_{\text{orth}}$ that compounds the computational cost for large systems.

\par 
In this work, we present G(Wa)NN, an $E(3)$-equivariant graph neural network designed to directly generate the spin-orbit coupled electronic Hamiltonians of solids in a Wannier basis. Because the Wannier basis is naturally orthogonal and highly localized, the resulting Hamiltonian matrix remains strictly sparse. This sparsity enables rapid inference and generation of electronic structures for macro-scale systems. Furthermore, utilizing a Wannier representation allows us to exploit subspace projection. For instance, while a large cell may contain hundreds of deep, fully occupied bands that bloat the Hilbert space, transport properties are governed entirely by states near the Fermi energy. G(Wa)NN natively accommodates the construction of a reduced Hilbert space focused solely on this relevant energy window, dramatically accelerating downstream transport calculations\cite{Pizzi2020,tiwari2026meta}.
\par
To make this framework immediately accessible, we also introduce \textit{Tailwater}, a Python package providing a streamlined API interface to G(Wa)NN and an optimized post-processing suite. \textit{Tailwater} automates the projection of Hamiltonians generated by G(Wa)NN into custom low-energy windows—directly emulating Wannierization workflows familiar to users of Wannier90—while preserving full basis information. Additionally, the ecosystem includes a suite of KPM-based functions designed to leverage this orthogonal, sparse format for true linear-scaled calculation of spectral observables, alongside workflows to locally finetune the base G(Wa)NN model on user-provided DFT datasets.

\par
In the remainder of this work we provide details of the training set, the model architecture, performance benchmarks, an overview of the finetuning capabilities, and the Tailwater python package which serves as a user-friendly front-end.

\begin{figure}
    \centering
    \includegraphics[width=8cm]{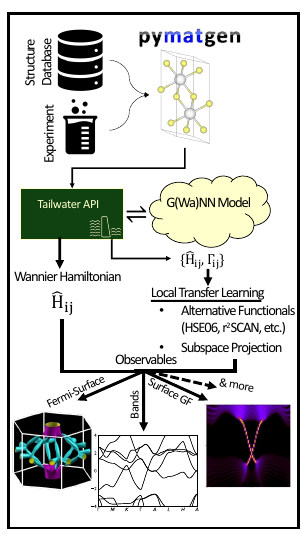}
    \caption{\textbf{Schematic overview of the Tailwater ecosystem:} Crystal structures obtained from experimental characterization or high-throughput structural databases are processed via pymatgen and fed directly into the Tailwater API. The API interfaces with the core $E(3)$-equivariant G(Wa)NN model to generate the gauge-standardized, orthogonal Wannier Hamiltonian ($\hat{H}_{ij}$). The API also yields the graph embedding ($\Gamma_{ij}$) to drive local transfer learning workflows, enabling efficient finetuning on alternative exchange-correlation functionals (e.g., $\text{HSE06}$, $\text{r}^2\text{SCAN}$) or immediate subspace projection into low-energy windows. The predicted sparse, orthogonal Hamiltonian serves as a high-performance engine for computing downstream physical observables—including Fermi surfaces, electronic band structures, surface Green's functions (Surface GF), and large-scale transport properties—bypassing the traditional scaling bottlenecks of first-principles methods.}
    \label{fig:Fig1}
\end{figure}

\par

\begin{figure*}
    \centering
    \includegraphics[width=18cm]{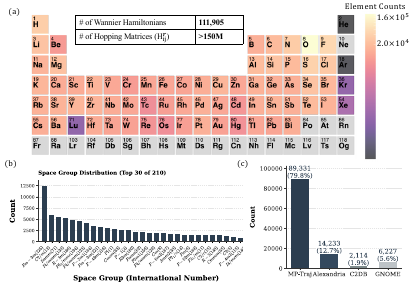}
    \caption{\textbf{Composition and structural diversity of the training dataset:} (a) Periodic table distribution highlighting the elemental diversity of the training set, which spans 69 elements and totals 111,905 Wannier Hamiltonians comprising over 150 million distinct spatial hopping matrices ($\hat{H}_{ij}^{\mathbf{R}}$). (b) Structural diversity across space groups, showing the distribution of the top 30 most frequent space groups (out of 210 total represented in the dataset), led by the high-symmetry cubic $Fm\bar{3}m$ (225) phase. (c) Breakdown of data provenance across various source materials repositories, illustrating that the majority of the training data originates from the Materials Project Trajectories (MP-Traj), supplemented by the Alexandria, C2DB, and GNOME databases.}
    \label{fig:Dataset}
\end{figure*}
\section{The Dataset} 
Progress in machine learning for electronic structure has been fundamentally bottlenecked by a severe dearth of data and the stringent physical constraints required to generate it. In modern DFT software (e.g., Quantum ESPRESSO\cite{QE-2020,QE-2017,QE-2009}, VASP\cite{hafner2008ab}, SIESTA\cite{soler2002siesta}), electronic details are contained within large wavefunction files that are routinely discarded after computing macroscopic properties to conserve disk space. Consequently, while public databases like the Materials Project\cite{jain2013commentary} and OQMD\cite{saal2013materials} contain millions of computed energies, forces, and stresses, they do not harbor the fundamental wavefunctions or Hamiltonians. The success of machine learning interatomic potentials (MLIPs) was enabled by these vast databases because energies and forces are basis-independent and trivial to standardize\cite{barros2026open}. ML Hamiltonians, however, face a uniquely rigorous barrier: gauge covariance. Because wavefunctions and Hamiltonians are gauge-dependent, ad-hoc data aggregation is impossible; learning the electronic structure dictates that all training data must be generated with systematic consistency. 
\par 
In this work we pursue transformation of the electronic structure to a Wannier basis via the Wannier90 software package\cite{Pizzi2020} for training of the ML model. This transformation offers numerous key benefits, most important of which is the resulting orthogonal basis. As stated in the introduction, the goal of this work is progress towards device relevant electronic structure computations. This goal can be restated as computation of transport quantities at the million-atom limit. From the perspective of the author, this will require a workflow which utilizes sparse matrix operations throughout, and likely relies on linear scaling methods (e.g. KPM) for determination of physical observables. While, deep generation of Wannier models introduces new challenges due to additional gauge freedom, it is argued that the result of a Hamiltonian in an orthogonal basis warrants the additional effort. 
\par
Compounding our motivation to generate Hamiltonians in a Wannier basis is the large ecosystem of software built for Wannier Hamiltonians, allowing for seamless integration into discovery workflows. These benefits have been exploited in prior machine-learning studies, such as Ref.~\cite{qi2025bridging} where the authors create dense networks to learn individual hopping parameters for permutations of a small bilayer system, demonstrating transferability to much larger twisted bilayer structure. This study demonstrated a proof-of-concept for machine learning of Wannier Hamiltonians in the context of a single stoichiometry. We aim to extend this idea to the level of a foundational model. 
\par 
To this end, we construct a dataset composed of more than 111K Wannier Hamiltonians generated for structures from the MP-Traj dataset\cite{deng_2023_chgnet,jain2013commentary}, Computational two-dimensional materials database (C2DB)\cite{haastrup2018computational}, GNOME\cite{merchant2023scaling}, and Alexandria\cite{ghahremanpour2018alexandria}. This represents the single largest dataset of Wannier Hamiltonians, to the author's knowledge. The protocol to create the dataset is as follows:
\par 
\emph{Structure selection: } Atomic structures are selected excluding (I) systems with more than 20 atoms in the primitive unit cell, (II) structures containing lanthanides or elements with atomic number greater than 86, and (III) magnetic systems. 
\par 
\emph{First-principles computations: }
For each of the resulting compounds we begin with density functional theory computations via the Quantum Espresso software package \cite{QE-2020,Perdew1996}. We utilize fully relativistic norm-conserving pseudopotentials from the PseudoDojo library~\cite{van2018pseudodojo}  along with a maximal plane-wave energy cutoff of 60\,Ry. All computations incorporate spin-orbit coupling (SOC). The Brillouin zone is sampled using a 6 $\times$ 6 $\times$ 6 Monkhorst-Pack $k$-point grid \cite{monkhorst1976special}. This grid is sufficient for convergence of the DFT computations for the smallest unit cell considered in this work. The choice of a uniform grid is made to prioritize the creation of high-quality Wannier tight-binding models rather than maximize computational efficiency.
\par 
It is important to comment further on our choice to utilize structures from the MP-Traj dataset. Unlike structures selected from other data-sources, out-of-equilibrium structures are included among those selected from MP-Traj. This is a dataset commonly used for training MLiPs, and the out-of-equilibrium structures are generated in the relaxation process. For each relaxation trajectory, we selected frames only where the minimum bond length difference was greater than $0.05\AA$, and placed a cap of 5 frames per material. This was done to avoid inclusion of redundant near-equilibrium structures. Ultimately, the MP-Traj contribution to the dataset is 89,331 structures from 34,564 materials. The purpose for including such Hamiltonians is to make progress towards capturing finite-temperature effects as well as improved performance at surfaces and interfaces where reconstruction can occur. 
\par 
\emph{Wannier Hamiltonian Generation: }
Upon completion of a self-consistent and non self-consistent DFT computation, a Wannier Hamiltonian is generated using the Wannier90 software package\cite{Pizzi2020}. 
One reason for the lack of a single large database of Wannier Hamiltonians despite their utility is that the process of generating a Wannier Hamiltonian from first-principles computations can be computationally expensive and there exists no guarantee that the basis conversion will yield accurate results. The outcome must be verified by benchmarking against the DFT computations. The crux of this procedure involves selecting an orbital basis for the Wannier Hamiltonian as well as selecting a subset of bands for which Wannier90 should attempt to fit the Hamiltonian using the selected basis. While methods for automated creation of Wannier Hamiltonians exist, we have not employed such methods in this work due to the need for a fixed basis for each element to create uniformity among the dataset for training of the model. 
\par 
We select the basis shown in Tab.~\ref{tab:basis}, creating a uniform number of orbitals for each element across all computations. In each Wannier90 model, the number of Wannier functions is exactly double the sum of the number of orbitals for each element, the factor of two is present to account for SOC. In the DFT computations we ensure the total number of states is equal to the number of Wannier functions plus ten additional states. The additional ten states are included such that the number of bands is greater than the number of Wannier functions, allowing for a disentanglement procedure within Wannier90 followed by maximal localization of the Wannier centers through an iterative process. While disentanglement has the potential to introduce noise into the dataset, our motivation for including this step is two-fold. First, we select the frozen window to include \emph{all} occupied bands and terminate at $1eV$ above the Fermi energy. This is primary energy window of interest for all relevant physical computations and the disentanglement procedure increases accuracy in this window. Second, the maximal localization procedure is beneficial as it has been found to act in a manner similar to gauge fixing, reducing a key potential source of noise.

\begin{table}[b]
\caption{\sffamily \textbf{G(Wa)NN universal basis set.} In training and for the initial forward pass, G(Wa)NN uses the following orbital basis. 
}
\setlength{\tabcolsep}{5pt}
\begin{ruledtabular}
\begin{tabular}{lccc}
 $Z$ & $s^{\uparrow\downarrow}$ & $p^{\uparrow\downarrow}$ & $d^{\uparrow\downarrow}$ \\
\hline
 $Z\leq 2$     & \checkmark &            &            \\
 $2<Z\leq 18$  & \checkmark & \checkmark &            \\
 $18<Z$        & \checkmark & \checkmark & \checkmark \\
\end{tabular}
\end{ruledtabular}
\label{tab:basis}
\end{table}

\par 
Once the Wannier Hamiltonian is created, two primary checks are performed, (I) the band structure is compared to the DFT band structure at a random selection of points in reciprocal space, if the mean average error between the DFT and Wannier band structure is greater than $0.05eV$ the model is rejected and (II) the spread of all Wannier centers must be $<1.5 \times |\mathbf{v}_{max}|$ where $\mathbf{v}_{max}$ is the largest magnitude lattice vector. This is similar to the requirements listed in Ref.~\cite{fang2025dataset}. Statistics for all generated Wannier Hamiltonians passing these requirements are shown in Fig.~\eqref{fig:Dataset}. 

\section{G(Wa)NN Model and Training}
We summarize the mathematical structure of our model, an equivariant message-passing graph neural network that predicts the electronic Hamiltonian, $\Ham$, of a material directly from its atomic geometry. The network follows the operator-learning formalism in which the target is an operator- valued quantity expressed in a localized orbital basis, and is built from $SO(3)$ tensor products coupled through Clebsch–Gordan coefficients. 
\par 

In a basis of localized orbitals $\{\ket{\varphi_i}\}$ that transform like
spherical harmonics $Y^{\ell}_{m}(\hat{\mathbf{r}})$ under rotation, the
single-particle electronic structure problem takes the form of a
generalized eigenvalue problem,
\begin{equation}
\begin{aligned}
  \Ham\,\bm{\psi} &= \varepsilon\,\Smat\,\bm{\psi}, \\
  H_{ij} &= \braket{\varphi_i | \hat{H}(\mathbf{r}) | \varphi_j}, \\
  S_{ij} &= \braket{\varphi_i | \varphi_j},
\end{aligned}
\end{equation}
where $\hat{H}(\mathbf{r})$ is the effective Hamiltonian operator. Our model
learns the mapping from an atomic structure to the matrix $\Ham$ expressed
in a localized \emph{Wannier} basis, predicting both the on-site blocks
$\Ham_{ii}$ (per atom / graph node) and the inter-site blocks
$\Ham_{ij}(\mathbf{R})$ (per directed edge between atom $i$ and a
periodic image of atom $j$). By working in a Wannier basis we can define
$\Smat=\mathbb{1}$, the identity matrix.

The basis is \emph{spinful}: each block carries a two-component spin
structure, so that a block coupling the on-site $s\oplus p\oplus d$
manifold ($9$ spatial orbitals) to itself is a $18\times 18$ complex
matrix,
\begin{equation}
\begin{aligned}
  \Ham^{\text{block}} &\in \mathbb{C}^{(9\cdot 2)\times(9\cdot 2)}, \\
  18 &= \underbrace{(2\cdot0{+}1)+(2\cdot1{+}1)+(2\cdot2{+}1)}_{\text{spatial}\,=\,9}
        \times \underbrace{2}_{\text{spin}} .
\end{aligned}
\end{equation}
Each spatial sub-block indexed by orbital degrees $(\ell_\alpha,\ell_\beta)$
on the two centres,
\begin{equation}
  \Ham^{\alpha\beta}_{ij}(m_\alpha,m_\beta)
   = \braket{\ell_\alpha, m_\alpha | \hat{H} | \ell_\beta, m_\beta},
\end{equation}
is, before the spin structure is attached, an object that transforms as the
tensor product $\irr{\ell_\alpha}{\ell_\beta}$, of dimension
$(2\ell_\alpha+1)\times(2\ell_\beta+1)$.

\subsection{Equivariant features and Clebsch--Gordan tensor products}

Node and edge features are organized as collections of spherical tensors
(\emph{irreps}) indexed by an angular degree $\ell\le \ell_{\max}$ and a
parity, replicated over a number of learnable channels. Under a rotation
$R\in\mathrm{SO}(3)$ a degree-$\ell$ feature transforms by the Wigner-$D$
matrix,
\begin{equation}
\begin{aligned}
  \mathbf{x}^{(\ell)} &\longmapsto \Dmat{\ell}(R)\,\mathbf{x}^{(\ell)}, \\
  \Dmat{\ell}(R) &\in \mathbb{R}^{(2\ell+1)\times(2\ell+1)},
\end{aligned}
\end{equation}
and a feature spanning several degrees transforms under the direct sum
$\bigoplus_{\ell} \Dmat{\ell}(R)$.

The fundamental equivariant operation is the coupling of two spherical
tensors of degrees $\ell_1$ and $\ell_2$ into an output of degree $L$. The
tensor product decomposes into irreducible representations,
\begin{equation}
  \ell_1\otimes\ell_2
    = \bigoplus_{L=\lvert \ell_1-\ell_2\rvert}^{\,\ell_1+\ell_2} L,
\end{equation}
with components obtained through the Clebsch--Gordan (CG) coefficients
$\CG{L,M}_{(\ell_1,m_1)(\ell_2,m_2)}$,
\begin{multline}
  \bigl(\mathbf{x}^{(\ell_1)}\otimes\mathbf{y}^{(\ell_2)}\bigr)^{(L)}_{M} = \\
  \sum_{m_1=-\ell_1}^{+\ell_1}\ \sum_{m_2=-\ell_2}^{+\ell_2}
  \CG{L,M}_{(\ell_1,m_1)(\ell_2,m_2)}\,
  x^{(\ell_1)}_{m_1}\, y^{(\ell_2)}_{m_2}.
  \label{eq:cg}
\end{multline}
These couplings are realized as weighted tensor products in which the
weight assigned to each $(\ell_1,\ell_2,L)$ path is produced by a radial
network. In our model the radial weights are conditioned not only on the
interatomic distance $\lvert\mathbf{R}_{ij}\rvert$ but also on the chemical
identities $(Z_i,Z_j)$ of the two endpoints, so that chemically distinct
bonds at equal length receive distinct couplings. The model retains the
full $\mathrm{SO}(3)$ Clebsch--Gordan coupling of Eq.~\eqref{eq:cg} (with
parity tracked, i.e.\ $\mathrm{O}(3)$) rather than reducing it to an
$\mathrm{SO}(2)$ form. All operations on the graph utilize the e3nn software
package~\cite{geiger2022e3nn}.

\subsection{Hamiltonian reconstruction}

The final per-node and per-edge embeddings are mapped to Hamiltonian blocks
by inverting the decomposition of Eq.~\eqref{eq:cg}: the network emits a set
of covariant coefficients that are recombined with a precomputed basis of
coupling tensors. For the spatial part, the $(\ell_\alpha,\ell_\beta)$
sub-block is assembled from the same CG coefficients,
\begin{multline}
  \Ham^{\alpha\beta}(m_\alpha,m_\beta) = \\
  \sum_{L=\lvert\ell_\alpha-\ell_\beta\rvert}^{\ell_\alpha+\ell_\beta}\
  \sum_{M=-L}^{+L}
  \hat{h}^{(L)}_{M}\,
  \CG{L,M}_{(\ell_\alpha,m_\alpha)(\ell_\beta,m_\beta)},
\end{multline}
where $\hat{h}^{(L)}_{M}$ are the covariant network outputs and the CG
coefficients are related to Wigner-$3j$ symbols in the standard
way~\cite{varshalovich}.

The spin structure is attached by coupling the spatial tensors to the
two-dimensional spin space. The block basis is the direct sum of a
spin-scalar channel and a spin-vector channel,
\begin{align}
  \mathbf{C}^{(L)}_{\mathrm{scalar}}
    &= \bigl(\mathbf{C}^{(L)}_{\mathrm{spat}}
       \otimes \mathbb{1}_{2}\bigr)^{(L)},
    \label{eq:spin-scalar}\\
  \mathbf{C}^{(L_{\mathrm{out}})}_{\mathrm{vector}}
    &= \bigl(\mathbf{C}^{(L)}_{\mathrm{spat}}
       \otimes \boldsymbol{\sigma}\bigr)^{(L_{\mathrm{out}})},
    \qquad |L-1| \le L_{\mathrm{out}} \le L+1,
    \label{eq:spin-vector}
\end{align}
where $\mathbb{1}$ and
$\boldsymbol{\sigma} = (\sigma_{x}, \sigma_{y}, \sigma_{z})$ are
treated as the spin-scalar ($\ell = 0$) and even-parity spin-vector
($\ell = 1$) objects, respectively, and
$(\,\cdot \otimes \cdot\,)^{(L)}$ denotes the coupled product of
Eq. (6), evaluated component-wise with products of
components understood as Kronecker products into the spin space. The
predicted block is the covariant combination
$\mathbf{H}^{\mathrm{block}} = \sum_{k} a_{k}\, \mathbf{C}_{k}$ of
these basis tensors with complex network coefficients $a_{k}$.
\par

\begin{figure*}
    \centering
    \includegraphics[width=15cm]{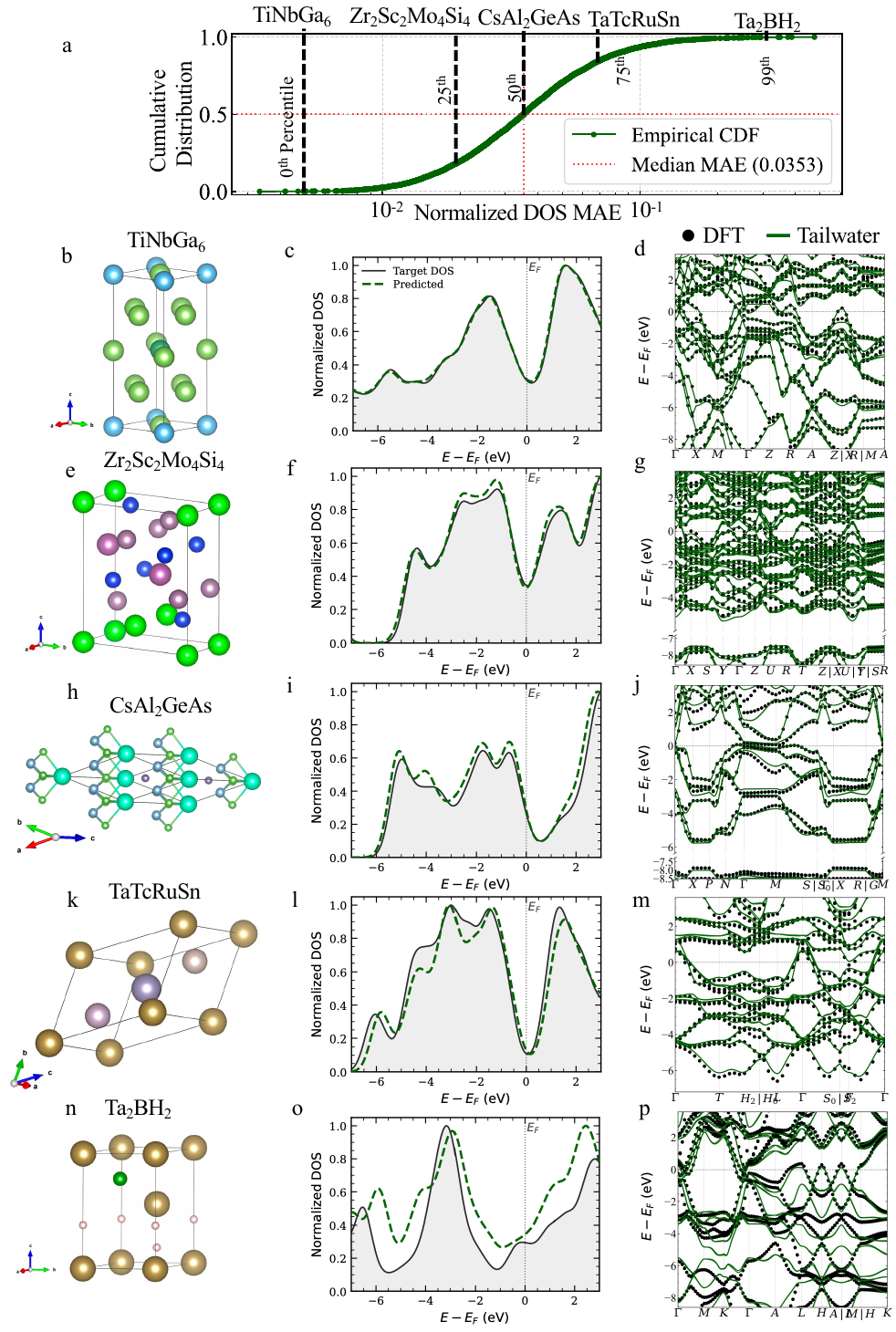}
    \caption{\textbf{Performance on validation set:} (a) Cumulative distribution of mean-average error (MAE) for the density of states (DOS) evaluated over the 2000 material validation set. The MAE for the DOS is determined by computing the density of states over the energy range $E\in[-7eV,3eV]$ on a discrete grid with spacing, $0.025eV$ via Eq.~\eqref{eq:DOS} using both the target and predicted Hamiltonian. The MAE is then computed via Eq.~\eqref{eq:rho_dos}. The median MAE of 0.035 demonstrates a high level of accuracy. (b)-(p) Structures are selected from the 0$^{th}$, 25$^{th}$, 50$^{th}$, 75$^{th}$ and 99$^{th}$ percentiles of the cumulative distribution shown in (a). These sample are representative of the best to worst materials in terms of DOS prediction respectively. For each material, from left to right, the crystal structure, predicted and target DOS, and predicted and target band structure are provided. }
    \label{fig:benchmark}
\end{figure*}
\par

\subsection{Model training and results}
The model is trained against a composite objective combining a real-space
and a momentum-space term,
\begin{equation}
    \mathcal{L}_{total}=\mathcal{L}_{R}+\mathcal{L}_{k},
\end{equation}
where $\mathcal{L}_{R}$ is the mean absolute error on the predicted
Hamiltonian matrix elements,
\begin{equation}
\mathcal{L}_{R}= \lVert H_{pred}-H_{target}\rVert .
\end{equation}

\subsubsection{Momentum-space loss}
The momentum-space term $\mathcal{L}_{k}$ acts on the per-$\mathbf{k}$ eigenvalues $\epsilon_n^{p}$ and $\epsilon_n^{t}$ of the predicted and target Bloch Hamiltonians ($n$ the band index, the $\mathbf{k}$ dependence implicit) and combines three complementary terms.

\paragraph{Band-edge--anchored eigenvalue error.}
A weighted eigenvalue MAE ($\mathcal{L}_{\text{eig}}$) that assigns full weight to states near the valence- and conduction-band edges and smoothly down-weights states far from the Fermi level, concentrating accuracy where band gaps, effective masses, and transport properties are determined.

\paragraph{Level spacing.}
An error on band-to-band spacings ($\mathcal{L}_{\Delta}$) that is invariant to global energy offsets, constraining the relative positions of neighboring bands rather than their absolute energies.

\paragraph{Crossing protection.}
A spacing term ($\mathcal{L}_{\times}$) concentrated on near-degenerate band pairs, discouraging artificial gapping of band crossings required by the underlying symmetries.

\paragraph{Combined loss.}
These combine as
\begin{equation}
\mathcal{L}_{k} = \mathcal{L}_{\text{eig}}
            + \lambda_{\Delta}\,\mathcal{L}_{\Delta}
            + \lambda_{\times}\,\mathcal{L}_{\times},
\label{eq:eigval_total}
\end{equation}
with $\lambda_{\Delta}$ and $\lambda_{\times}$ set so that the spacing and crossing terms refine rather than overwhelm the eigenvalue error.

\section{Model Benchmarks}
It is common to report loss metrics based on Hamiltonian matrix elements\cite{zhong2026universal,qian2026equivariant,qian2026equivariant}, (the network achieves a MAE of 10meV on the validation set), however such metrics can be misleading due to gauge co-variance of the target Hamiltonian. We thus focus on gauge-invariant observables. This allows for comparison to existing machine learning approaches and overcomes underlying gauge-noise. 

\subsection{Density of states and eigenspectra}
The density of states (DOS) has been a common target for past deep generative approaches as it represents a mapping from a given crystal structure and energy level, $E$, to a scalar $\rho(E)$. Furthermore, the density of states is of general use to the scientific community and has been collected in a number of databases including the Materials Project, NOMAD\cite{sbailo2022nomad} and C2DB. Recently, a multimodal\cite{lee2023density} and point-edge-transformer (PET)\cite{how2026universal} approach have yielded impressive accuracy in prediction of the density of states. These references further utilize the density of states to extract the band gap. This provides a useful comparison to G(Wa)NN as models which are not designed for explicit prediction of band-gap but from which it can be extracted. 
\par 
Here we follow the convention used in Ref.~\cite{lee2023density}, computing the density of states for both the target and predicted Wannier Hamiltonian using the eigenvalues from a $10 \times 10 \times 10$ grid of $\mathbf{k}$-points as, 
\begin{equation}\label{eq:DOS}
    \rho(E)=\frac{1}{N_k} \sum_{k} \sum_{n} \frac{1}{\sigma \sqrt{2\pi}} \exp\left( -\frac{1}{2} \left( \frac{E - E_{n,k}}{\sigma} \right)^2 \right),
\end{equation}
where $N_{k}$ is the number of $k$-points and $E_{n,k}$ is the eigenvalue of band $n$ at a given point in the Brillouin zone, $k$. This is computed for $E\in[-7eV,3eV]$ on a discrete grid with spacing, $0.025eV$. The resulting curve is then normalized by its maximum value such that the DOS values fall exclusively between 0 and 1. The MAE is then computed as, 
\begin{equation}\label{eq:rho_dos}
    MAE_{\rho}=\frac{1}{N_{E}}\sum|\hat{\rho}_{pred}(E)-\hat{\rho}_{target}(E)|,
\end{equation}
where the sum is over all discrete points on the DOS curve, $N_{E}$, and $\hat{\rho}$ is the normalized DOS. 
\par 
In Fig.~\eqref{fig:benchmark}(a) the cumulative distribution of the MAE over a validation set of 2000 materials from the GNOME dataset is shown. The model achieves an overall MAE of 0.04, comparable with that of Ref.~\cite{lee2023density}, which is stated as 0.08 for in-distribution systems and 0.122 for out-of-distribution crystals. We emphasize that this work uses a distinct test and train set to Ref.~\cite{lee2023density} making direct comparison impossible. To provide a qualitative understanding of the accuracy, the crystal structure, density of states, and band structure for example systems falling in the $0,25,50,75,$ and $99^{th}$ percentile of the cumulative distribution function are plotted in Fig.~\eqref{fig:benchmark}(b)-(p). We note that in nearly all cases accuracy is greatest near the Fermi energy, reflecting the loss function design which prioritizes accuracy in the low-energy sector. This allows for even systems falling in the worst $25\%$ of predictions to provide physically useful information. We also expect enhanced noise in the conduction sector due to the disentanglement procedure in the Wannierization process itself.

\subsection{Band gap}
\begin{figure}
    \centering
    \includegraphics[width=8cm]{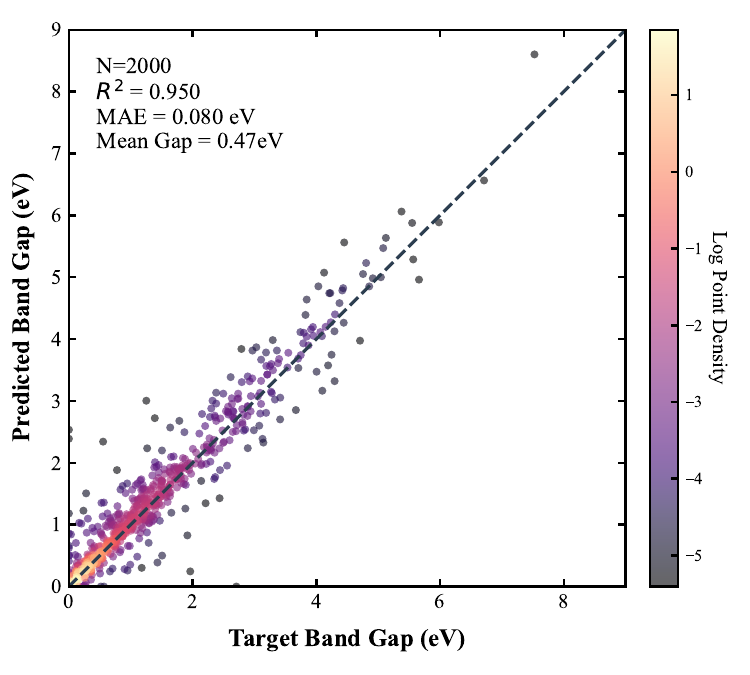}
    \caption{\textbf{Band Gap Prediction on Validation Set:} Parity plot detailing band-gap prediction for 2000 materials in the validation set (not seen in training), selected from the GNOME database.}
    \label{fig:BandGap}
\end{figure}

Aside from being a useful physical quantity, prediction of band gap serves as a convenient metric for assessing the quality of the model as we can compare to the wide range of existing machine learning models constructed for this purpose. Furthermore, band gap prediction is a proxy for the quality of the predicted eigen-spectra near the Fermi energy. We test the band-gap prediction of the model against the same validation set of 2000 materials from the GNOME dataset analyzed in Fig.~\eqref{fig:benchmark}. The mean band gap among materials in this set is $0.47eV$. The results of the Tailwater model are shown in Fig.~\eqref{fig:BandGap}, demonstrating a mean average error of 0.08$eV$. To provide a comparison, we have trained a crystal graph convolutional neural network (CGCNN)\cite{cgcnn}, atomistic line graph neural network (ALIGNN)\cite{choudhary2021atomistic}, and connectivity optimized nested line graph network (coGN)\cite{ruff2024connectivity} on the same set of training data. Training was done using the default settings for all models with the exception of the CGCNN for which the graph radius was increased to $16\AA$. The results are shown in Tab.~\eqref{tab:Bandgap}.

\begin{table}[b]
\caption{\sffamily \textbf{Benchmarking band-gap predictions} Band-gap MAE and $R^{2}$ for 2000- material validation set from GNOME\cite{merchant2023scaling} across models trained using the same dataset. Best values are shown in bold text.
}
\setlength{\tabcolsep}{5pt}
\begin{ruledtabular}
\begin{tabular}{lccc}
 Model & MAE (eV) & $R^{2}$ \\
\hline
 G(Wa)NN     & \textbf{0.080} &    \textbf{0.95}                 \\
 CGCNN\cite{cgcnn}  & 0.303 & 0.72    \\
 ALIGNN\cite{choudhary2021atomistic}        & 0.092 & 0.93  \\
 coGN\cite{ruff2024connectivity}        & 0.111 & 0.91 \\
\end{tabular}
\end{ruledtabular}
\label{tab:Bandgap}
\end{table}

The performance of CGCNN, ALIGNN, and coGN models are in-line with existing literature results\cite{choudhary2021atomistic,dunn2020benchmarking}. While our results can not be directly compared to the MatBench leader board\cite{dunn2020benchmarking}, we note that coGN and ALIGNN currently rank first and fourth respectively on the MatBench bandgap leaderboard. The relative improvement of our model indicates that G(Wa)NN is approaching or at the state-of-the-art for band-gap prediction. 

\begin{figure*}
    \centering
    \includegraphics[width=18cm]{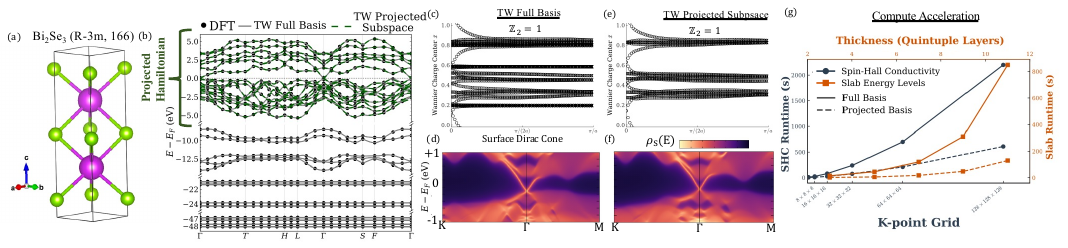}
\caption{\textbf{Example Workflow for $\mathbf{\text{Bi}_2\text{Se}_3}$: Topological analysis and computational acceleration benchmarks.} (a) Primitive unit cell of rhombohedral ($R\bar{3}m$) $\text{Bi}_2\text{Se}_3$. (b) Electronic band structure comparing ground-truth DFT (black circles) with the Tailwater (TW) model predicted full-basis Wannier Hamiltonian (gray lines) across all 90 bands, and the 30 band projected Wannier Hamiltonian (dashed green lines), showing excellent agreement. (c) Evolution of Wannier charge centers calculated via Z2Pack in the $k_y = 0$ plane for the TW full basis yielding $\mathbb{Z}_2 = 1$. (d) Surface spectral density computed using Tailwater’s highly optimized Kernel Polynomial Method (KPM) routines for the full basis resolving the gapless surface Dirac cone characteristic of the topological insulator phase. (e)-(f) The Wannier charge centers and surface spectral density computed for the TW projected low-energy subspace, revealing identical physics.  (g) Spin-Hall conductivity (SHC) runtime comparison using WannierBerri evaluated across increasingly dense $k$-point grids (dark gray lines) from $4\times4\times4$ up to $128\times128\times128$, demonstrating multi-orders-of-magnitude acceleration for the projected subspace. Benchmark of computational scaling for solving $\Gamma$-point slab eigenvalues as a function of quintuple layer thickness (orange lines), highlighting massive reduction in slab runtime afforded by projected basis set.}
\label{fig:Projection}
\end{figure*}
\section{The Tailwater interface}
In an effort to accelerate materials discovery efforts, the Tailwater (TW) python package has been developed to provide a simplified API interface for accessing the G(Wa)NN model, as well as provide a library of post-processing function that integrate seamlessly with the interface to aid construction of automated workflows. This package can be downloaded via PyPi and a full tutorial along with code examples are provided on Github\cite{tyner_github} and \url{https://tailwater.readthedocs.io}. We emphasize that, whenever possible, an implementation based on the kernel polynomial method (KPM) is provided. This package is open-source and compatible with any Wannier Hamiltonian, not only those produced by G(Wa)NN. Our aim is to provide a useful community tool for easily navigating the Wannier ecosystem.  
\par 
A schematic detailing the steps for accessing G(Wa)NN via the API and post-processing the Hamiltonian with the Tailwater front-end package is given in Fig.~\eqref{fig:Fig1}. We note that the workflow requires only a PyMatgen\cite{ong2013python} structure as input. The resulting Wannier Hamiltonian is returned as a sparse matrix in $npz$ format. For small system sizes, this is automatically converted to an HDF5 file for immediate use by the Tailwater or TBModels\cite{z2pack} software package. Alternatively a built in command for writing the model to a wannier90\textunderscore hr.dat file is available. For large systems the Hamiltonian is left as a sparse matrix with built in optionality for specifying a reciprocal space vector. Optionality to load the Hamiltonian as a sparse matrix in Kwant\cite{groth2014kwant} or Pybinding\cite{moldovan2020pybinding} is also provided.

\begin{figure*}
    \centering
    \includegraphics[width=18cm]{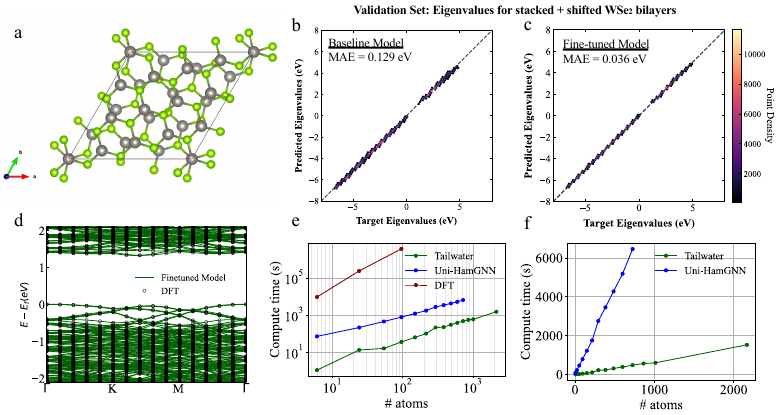}
\caption{\textbf{Finetuning for OOD systems:} (a) Primitive unit cell of twisted bilayer WSe$_{2}$ at $\theta=32.2^{\circ}$, containing 78 atoms. (b)-(c) Parity plot of DFT versus predicted eigenvalues for the (b) baseline and (c) finetuened model over a $4\times 4\times 1$ grid of $\mathbf{k}$-points. Eigenvalues are computed for the validation set of stacked and shifted WSe$_{2}$ bilayers. The baseline model demonstrates strong out-of-the-box accuracy but can be quickly improved through finetuning. (d) The DFT and predicted band structure for $\theta=32.2^{\circ}$ twisted bilayer WSe$_{2}$ using the fine-tuned model. (e)-(f) Comparison of time to generate the supercells of bilayer WSe$_{2}$ between the Uni-HamGNN foundational model and the Tailwater ecosystem on an Apple M3 Max. A further comparison is made in (e) with Quantum Espresso (DFT) on an HPC Cluster using an AMD EPYC 7713 processor.}
\label{fig:Finetune}
\end{figure*}

\subsection{Subspace projection of the Hamiltonian}
The use of Wannier Hamiltonians, rather than those derived from LCAO DFT, has proliferated in part due to the ability to project into a relevant subspace containing only the states near the Fermi energy. This is a powerful capability which allows for reduction of the Hilbert space size by an order or magnitude or more for large systems with many filled valence orbitals that produce deep lying bands with no impact on relevant transport quantities. By reducing the Hilbert-space size, downstream computations, particularly Brillouin zone integrals required for determination of conductivity tensors, are vastly expedited. Additionally, a number of software packages are able to utilize the reduced Hilbert space size for accelerated computation of electron-phonon coupling and strongly-correlated phenomena. 
\par 
While the G(Wa)NN model always returns a Wannier Hamiltonian in a predetermined basis, fixing the number of orbitals present for each atomic species, the Tailwater package offers built-in functionality for projecting the Wannier Hamiltonian into a relevant energetic subspace. This process requires only defining an energetic window containing the bands the user desires to preserve. The user then defines a value for $\sigma$, which determines the shape of a Gaussian which is peaked at the center of the energetic, $E_{c}$, window and decays as $e^{-(E-E_c)^2/\sigma^2}$. This Gaussian is used to control the weight applied to eigenvalues within the energetic window when fitting the projected Hamiltonian. This process occurs on the users local machine and its effect is shown in Fig. \eqref{fig:Projection}. Generation of a model and subsequent implementation of the subspace projection routine is exemplified in the following code:
\begin{lstlisting}[language=Python]
import tailwater
from tailwater import tw_api_call,
subspace_projection
paths = tw_api_call(structure, USER, PSWRD,
"./outputs", "my_mat", project=True)
subspace_projection(
    start_lr = 1e-3,
    end_lr = 1e-5,
    num_epochs = N_EPOCHS,
    decay_sigma = SIGMA,
    device = "cpu", #CUDA when GPU available
    save_path = "./projection_out",
    embed_path = paths["embeddings"], #Input graph from API
    hr_npz_path = paths['npz'] #Target Hamiltonian for projection )
\end{lstlisting}

We emphasize that this process is computationally inexpensive and can be accomplished on personal machine for most systems. The resulting Hamiltonian is saved along with a file containing information about the basis of the projected Hamiltonian.

\subsubsection{Example: Topological Insulator Bi$_{2}$Se$_{3}$}

To demonstrate the utility and accuracy of the G(Wa)NN-Tailwater ecosystem, Fig.~\eqref{fig:Projection} illustrates an example workflow using the rhombohedral ($R\bar{3}m$) topological insulator $\text{Bi}_2\text{Se}_3$—a prototypical material widely studied in community packages like WannierTools~\cite{wu2018wanniertools}. In $\text{Bi}_2\text{Se}_3$, the non-trivial topological phase ($\mathbb{Z}_2 = [1;0,0,0]$) is fundamentally driven by strong relativistic spin-orbit coupling (SOC) originating from the heavy Bi and Se atoms. Specifically, SOC induces a critical band inversion at the $\Gamma$ point between the Bi $6p$ and Se $4p$ orbitals. Without an accurate representation of these relativistic interactions, the material is erroneously predicted to be a trivial band insulator. Therefore, precisely capturing SOC within our ML model is essential not only to reproduce the inverted bulk gap, but also to resolve the characteristic topologically protected Dirac surface states. As shown in the electronic band structure comparison Fig.~\eqref{fig:Projection}(b), the electronic Hamiltonian predicted via the Tailwater API (gray solid lines) exhibits exceptional agreement across all 90 valence and conduction bands when benchmarked against the ground-truth plane-wave DFT calculation (black dots). Utilizing Tailwater's automated projection capabilities, the full basis is then cleanly disentangled into a low-energy subspace capturing the 30 bands closest to the Fermi energy ($E_F$), which physically isolates the relevant $\text{Bi}$ and $\text{Se}$ $p$-orbitals (dashed green lines). To verify that the underlying topological physics is seamlessly preserved under machine-learning inference and subsequent subspace reduction, we compute the $\mathbb{Z}_2$ topological invariant in the $k_y = 0$ plane using Z2Pack\cite{z2pack} for both the (c) full and projected (e) Tailwater Hamiltonians. While the total number of tracked Wannier charge centers differs due to the reduced count of occupied bands in the projected subspace, both representations yield the identical topological classification of $\mathbb{Z}_2 = 1$. The non-trivial topology further manifests in the surface spectral densities calculated via Tailwater’s KPM routines; both the (d) full basis and (f) the projected subspace successfully resolve the characteristic gapless Dirac cone at the $\bar{\Gamma}$-point. Finally, the computational advantage of subspace projection is quantified in panel (g). Evaluating the spin-Hall conductivity (SHC) using WannierBerri\cite{tsirkin2021high} across a series of increasingly dense $k$-point grids (gray lines) reveals a massive reduction in runtime for the projected system compared to the full basis. This acceleration is mirrored in real-space scaling benchmarks (orange lines), where the computational time required to solve the $\Gamma$-point eigenvalues of a slab geometry scales drastically better for the projected Hamiltonian as a function of quintuple layer thickness, confirming the framework's potential to bring macro-scale quantum transport within reach.

\subsection{Fine-Tuning}
A current challenge facing deep-models for electronic structure is the ease with which fine-tuning can be implemented. In recent years fine-tuning of foundational MLiPs has become commonplace and shown to significantly improve accuracy in out-of-distribution (OOD) scenarios. Unlike MLiPs, fine-tuning deep-generative models for electronic structure often requires familiarity with the underlying LCAO DFT code and use of a basis set consistent with that used in generation of the training data. This poses a challenge given the diverse ecosystem of ab initio codes utilized by the community.  
\par 
To increase the utility of G(Wa)NN, particularly for systems requiring increased accuracy or OOD scenarios, the Tailwater package includes capabilities for fine-tuning using Wannier Hamiltonians created from any DFT code and with any basis set. Moire materials are an example target space where fine-tuning is useful to resolve low-energy electronic features\cite{Andrei2021,Cao2018,Cao2018b,Ghiotto2021,Mak2022,ZhaoS2023,Pixley2025}. Moire materials have attracted significant interest in recent years and been the target of many deep-generative models for electronic structure as they host flat-bands the primitive unit cell generally contains hundreds to thousands of atoms, making traditional first-principles methods too computationally expensive\cite{PhysRevResearch.4.043224,bao2024deep,yang2024identification,kaplan2025machine}.
\par 
Rather than a direct computation utilizing the full primitive unit cell, the typical approach for training a model to reproduce the electronic structure of a twisted bilayer is the so-called "stack and shift" procedure by which a bilayer unit cell or small supercell is created and one layer is systematically shifted over a grid. For each shift the Hamiltonian is generated and added to a database. The resulting database is then used to train a neural network which can capture the local inter and intra-layer matrix elements for the full twisted bilayer\cite{Fang2016,Carr2019,carr2020electronic,qi2025bridging,gong2023general,zhong2023transferable}.
\par 
Importantly, when carrying out the stack and shift method, it is common to retain only the relevant orbitals near the Fermi energy. For example, we create a dataset of stacked and shifted homo-bilayer WSe$_{2}$ Wannier Hamiltonians using Quantum Espresso. For this dataset computations are performed including the effects of spin-orbit coupling; to accurately account for the long-range van der Waals forces between the two layers, dispersion interactions were modeled using the DFT-D3 method\cite{grimme2010consistent}. It should be noted that the DFT-D3 method was not used in the generating data points within the initial training set; necessitating the fine-tuning procedure. For creation of Wannier Hamiltonians, only the $p$-orbitals of Se and the $d$-orbitals of W are included. This is distinct from the basis set for W and Se included in the initial training set, which contains $s,p,$ and $d$ orbitals. The Tailwater package handles this automatically, masking the orbitals that do not appear in the fine-tuning set when generating the Hamiltonian and computing the loss. We carry out training with a 90:10 train to validation split to finetune the model for analysis of twisted bilayer WSe$_{2}$, as shown in Fig.~\eqref{fig:Finetune}(a). A comparison of the DFT and G(Wa)NN generated eigenvalues for the validation set on a $4\times4\times 1$ grid of $\mathbf{k}$-points is shown for the baseline model (considering only the target energetic subspace) and fine-tuned model in Fig.~\eqref{fig:Finetune}(b)-(c) demonstrating reasonable baseline accuracy while also underscoring that a short fine-tuning can greatly enhance accuracy on out-of-distribution systems. It should be noted that fine-tuning, like subspace-projection is a computationally inexpensive local computation as it generally requires only the model head containing $\sim 61K$ parameters. As a local operation it also does not require sharing data through the API. 
\par 
Using the finetuned model the band structure is generated for $\theta=32.2^{\circ}$ twisted bilayer WSe$_{2}$ in Fig.~\eqref{fig:Finetune}(d) and compared to a direct computation in Quantum Espresso including spin-orbit coupling, a cutoff of $60Ry$, and the DFT-D3 method\cite{grimme2010consistent}. The comparison shows high-level agreement, underscoring the utility and accuracy of the model in out-of-distribution scenarios.  
\subsection{Handling extreme system sizes}
A primary goal of this work is to make transport and electronic structure computations of large systems (1K-10K+ atoms) accessible. This is an important step to accelerate device optimization in areas such as quantum computing, semiconductor manufacturing, and more. These computations are generally too expensive for plane-wave DFT, necessitating the use of alternative methods or semi-classical approximations. 
\par 
In order to accommodate extremely large system sizes, the inference pipeline has been optimized to construct and deliver the Hamiltonian in a sparse matrix format. This sparse matrix can be evaluated directly to extract eigenvalues or the Hamiltonian at arbitrary points in momentum space using the Tailwater package. Furthermore, the Tailwater package offers conversion functions to load the Hamiltonian as a sparse matrix in PyBinding\cite{moldovan2020pybinding} and Kwant\cite{groth2014kwant}. In systems with thousands of atoms remaining in a sparse matrix format is critical as a dense matrix representation can easily cause RAM requirements to induce computation failures. 
\par
Along with optimizing memory constraints, we focus on inference speed. The space of known and AI generated materials is rapidly expanding, exposing a need for electronic structure methods that can keep up. Ideally, these methods should operate in real-time to as tools for AI agents. The Tailwater interface optimizes speed by eliminating the need for any pre- or post-processing using DFT codes. An evaluation of the inference speeds for generation of the Hamiltonian is shown for increasingly large supercells of bilayer WSe$_{2}$ in Fig.~\eqref{fig:Finetune}(e)-(f). The comparison is made between the Tailwater and Uni-HamGNN\cite{zhong2026universal}, an existing foundational model for non-orthogonal Hamiltonians. The full inference pipeline for both models is executed on an Apple M3 Max. A further comparison is made with a direct computation using Quantum Espresso on an HPC Cluster using an AMD EPYC 7713 processor.

\subsection{Further capabilities and connection to the Wannier ecosystem}
The goal of the Tailwater python package is a user-friendly front-end that allows for seamless integration of the G(Wa)NN network into existing workflows. For that reason it includes a number of features that we briefly list here:
\begin{itemize}
    \item Band structure generation, automatic or on a custom $k$-path
    \item Density of states computed via KPM
    \item Surface spectral density on arbitrary surfaces computed with KPM or the Lopez-Sancho recursive Greens function
    \item Conversion tools to load the model in varying software packages such as: PythTB\cite{yusufaly2013tight}, Kwant,\cite{groth2014kwant} PyBinding\cite{moldovan2020pybinding}, WannierBerri\cite{tsirkin2021high}, and more. 
\end{itemize}
\par

\section{Outlook}

The principal obstacle to computational materials discovery is increasingly
one of scale rather than candidate availability: databases such as GNOME\cite{merchant2023scaling} now
enumerate vast numbers of structures, yet conventional first-principles
methods cannot evaluate their electronic and transport properties at
device-relevant sizes. The framework presented here is designed to address
this limitation directly. By predicting spin--orbit coupled electronic
Hamiltonians in an orthogonal, maximally localized Wannier basis, and
evaluating observables through linear-scaling Kernel Polynomial Method
routines, G(Wa)NN and the Tailwater interface represent progress towards a linear scaling alternative to the cubic scaling of
plane-wave DFT. To lower the barrier to adoption, a machine-readable prompt is provided at
\href{https://tailwater.io}{tailwater.io} that supplies coding agents (e.g.,
Claude Code, Codex) with the information required to operate the API and
Python package, so that a natural-language interface to the model is
immediately available.

\par
Several extensions are planned or already underway. These include expanding
the training corpus beyond 500{,}000 Wannier Hamiltonians to broaden chemical
and structural coverage; developing a native interface for electron--phonon
interactions to enable studies of transport, superconductivity, and
finite-temperature phenomena; incorporating hybrid functionals, principally
HSE06\cite{heyd2003hybrid}, to improve band-gap and spectral accuracy; and integrating magnetic
materials to extend the framework to spin-ordered systems.
\par
The model is presently accessible through the API, and the Tailwater package
is released as an open-source library compatible with Wannier Hamiltonians
from any source. We anticipate that this combination of a foundational model
and its associated post-processing infrastructure will provide a practical
basis for device-scale electronic-structure and quantum-transport studies.

\begin{acknowledgments}
A.T. would like to thank B. Baldassarri, J.M. Rondinelli, and F. Tyner for useful discussions. This research was supported in part by Lambda, Inc and participation in the NVIDIA Inception Program.

\end{acknowledgments}

\bibliography{Ref.bib}

\end{document}